\documentclass[11pt]{article}
\usepackage{graphics,amssymb}
\usepackage{epsfig}
\begin{document}

{
\setlength{\textwidth}{16.5cm}
\setlength{\textheight}{22.2cm}
\setlength{\hoffset}{-1.43cm}
\setlength{\voffset}{-.9in}

\thispagestyle{empty}
\renewcommand{\thefootnote}{\fnsymbol{footnote}}

\begin{flushright}
{\normalsize
SLAC-AP-127\\
July 2000}
\end{flushright}

\vspace{.8cm}

\begin{center}
{\bf\Large Analytical Formula for Weak Multi-Bunch Beam Break-Up in a Linac
\footnote{\small Work supported by
Department of Energy contract  DE--AC03--76SF00515.}}

\vspace{1cm}

{\large
Karl L.F. Bane and Zenghai Li\\
Stanford Linear Accelerator Center, Stanford University,
Stanford, CA  94309}

\end{center}
}
\vfill

\def\la{\langle} 
\def\ra{\rangle} 
\def\lm{\lambda}

\title{Analytical Formula for Weak Multi-Bunch Beam Break-Up in a Linac}
\author{Karl L.F. Bane and Zenghai Li}
\date{}
\maketitle

In designing linac structures for multi-bunch applications
we are often interested in estimating the effect of relatively
weak multi-bunch beam break-up (BBU), due to the somewhat complicated
wakefields of detuned structures.
This, for example, is the case for the
injector linacs of the JLC/NLC linear collider
project (see Ref.~\cite{BL}).
Deriving an analytical formula for such a problem is the subject
of this report.
Note that the more studied multi-bunch BBU
problem, {\it i.e.} the effect on a bunch train of a single strong
mode, the so-called ``cumulative beam break-up instability''
(see, {\it e.g.} Ref.~\cite{Lau}), is a 
somewhat different problem, and one for which the approach presented
here is probably not very useful.

In Ref.~\cite{chao} 
an analytical formula for {\it single-bunch} beam break-up 
in a smooth focusing linac, for the case without energy spread in
the beam,  is derived,
the so-called Chao-Richter-Yao (CRY) model for beam break-up. 
Suppose the beam is initially offset 
from the accelerator axis.
The beam break-up downstream is characterized by a strength parameter
$\Upsilon(t,s)$, where $t$ represents position within the bunch, and $s$
position along the linac.
When $\Upsilon(t,s)$ is small compared to 1, 
the growth in betatron amplitude in the linac is proportional to this parameter.
When applied to the special case of a uniform longitudinal
charge distribution, and a linearly growing wakefield, the result 
of the calculation becomes
especially simple. 
In this case the growth in orbit amplitude is given as an asymptotic
power series in $\Upsilon(t,s)$, and the series can
be summed to give a closed form, asymptotic
solution for single-bunch BBU.
The derivation of an analytic formula for {\it multi-bunch} BBU is almost 
a trivial modification of the CRY formalism.
We will here reproduce the important features of the 
single-bunch derivation of Ref.~\cite{chao}
(with slightly modified notation), and then
show how it can be modified to obtain a result applicable to multi-bunch BBU.

Let us consider the case of single-bunch beam break-up, where a beam is
initially offset by distance $y_0$ in a linac with 
acceleration and smooth focusing. 
We assume that there is no energy spread
within the beam.
The equation of motion is
\begin{equation}
{1\over E(s)}{d\over ds}
\left[E(s){dy(t,s)\over ds}\right]+ {y(t,s)\over\beta^2(s)}=
{e^2N_t\over E(s)}\int_{-\infty}^t dt^\prime\,y(t^\prime,s)\lambda_t(t^\prime)
W(t-t^\prime)\ ,\label{eqmotion}
\end{equation}
with $y(t,s)$ the bunch offset, a function of position within the bunch $t$,
and position along the linac $s$; with $E$ the beam energy, $[1/\beta(s)]$ the
betatron wave number, $eN_t$ the total bunch charge, $\lambda_t(t)$
the longitudinal charge distribution, and $W(t)$ the short-range 
dipole wakefield.
Our convention is that negative values of $t$ are toward the front of the 
bunch.
Let us, for the moment, limit ourselves to the problem of no acceleration and
$\beta$ a constant. 
A.~Chao in Ref.~\cite{chao} expands the solution to the equation of motion
for this problem in a perturbation series
\begin{equation}
y(t,s)=\sum_{n=0}^\infty y^{(n)}(t,s)\quad,\label{eqypert0}
\end{equation}
with the first term given by free betatron
oscillation [$y^{(0)}=y_0\cos (s/\beta)$].
He then shows that the solution for the higher terms
at position $s=L$, after many
betatron oscillations, is given by
\begin{equation}
y^{(n)}(t,L)\approx {y_0\over n!}\left(
{ie^2N_tL\beta\over 2E}
\right)^n R_n(t)e^{iL/\beta}\quad,\label{eqypert}
\end{equation}
with
\begin{eqnarray}
R_n(t)&=&\int_{-\infty}^t dt_1\,\lambda(t_1)W(t-t_1)
       \int_{-\infty}^{t_1} dt_2\,\lambda(t_2)W(t_1-t_2)\nonumber\\
      & &\cdots 
       \int_{-\infty}^{t_{n-1}} dt_n\,\lambda(t_n)W(t_{n-1}-t_n)\ ,
\label{eqrn}
\end{eqnarray}
and $R_0(z)=1$.
An observable $y$ is meant to be the real part of Eq.~\ref{eqypert0}.
The effects of adiabatic acceleration,
{\it i.e.} sufficiently slow acceleration so that the energy doubling
distance is large compared to the betatron wave length,
and $\beta$ not constant, can be added 
by simply replacing 
$(\beta/E)$ in Eq.~\ref{eqypert}
by $\langle\beta/E\rangle$, where
angle brackets indicate averaging along the
linac from $s=0$ to $s=L$.\footnote{Note that the terms $y_0 e^{iL/\beta}$
in Eq.~\ref{eqypert}, related to free betatron oscillation, 
also need to be modified in well-known ways
to reflect the dependence of $\beta$ on $E$.
It is the other terms, however,
which characterize BBU, in which we are interested.
}
For example, if the lattice is such that $\beta\sim E^\zeta$ then
$\langle\beta/E\rangle=(\beta_0/E_0)g(E_f/E_0,\zeta)$,
 where subscripts ``0'' and
``$f$'' signify, respectively, initial and final parameters, and
\begin{equation}
g(x,\zeta)= {1\over\zeta}\left({x^\zeta-1\over x-1}\right)
\quad\quad\quad[{\beta\sim E^\zeta}].
\end{equation}

Chao then shows that for certain simple combinations of bunch 
shape and wake function shape
the integrals in Eq.~\ref{eqrn} can
be performed analytically, and the result becomes  
an asymptotic series in powers of
a strength parameter.
For example, for the case of a uniform charge distribution of 
length $\ell$ (with the front of the bunch at $t=0$), and a wake
that varies as $W=W^\prime t$, the strength parameter is
\begin{equation}
\Upsilon(t,L)={e^2N_t LW^\prime t^2\beta_0\over 2E_0\ell}g(E_f/E_0,\zeta)
\quad.
\end{equation}
If $\Upsilon$ is small compared to 1, the growth is well approximated by 
$\Upsilon$.
 If $\Upsilon$ is large,
the sum over all terms can be performed to give
a closed form, asymptotic expression.

For {\it multi-bunch} BBU we are mainly concerned with the interaction of the
different bunches in the train, and will ignore wakefield forces
within bunches. The derivation is nearly identical to 
that for the single-bunch BBU.
However,
in the equation of motion, Eq.~\ref{eqmotion}, the independent variable $t$
is no longer a continuous variable,
but rather $t$ takes on discrete
values $t_m=m\Delta t$, where $m$ is a bunch index and $\Delta t$ is
the bunch spacing.
Also, $W$ now represents the long-range wakefield.
Let us assume that there are $M$, equally populated bunches in a train;
{\it i.e.} $N_t=MN$, with $N$ the particles per bunch.
The solution is again expanded in a perturbation series. In the solution,
Eq.~\ref{eqypert}, the $R_n(t)$, which are smooth functions
of $t$, are replaced by 
\begin{equation}
{\cal R}_m^{(n)}= {1\over M}\sum_{j=1}^{m-1} W[(m-j)\Delta t]{\cal R}_j^{(n-1)}
\quad,\label{eqcalr}
\end{equation}
(with ${\cal R}_j^0=1$), which is a function of a discrete 
parameter, the bunch index $m$. 
Note that ${\cal R}_m^{(1)}=S_m/M$, with $S_m$ the sum wake.

Generally the sums in Eq.~\ref{eqcalr} cannot be 
given in closed form, and therefore a closed, asymptotic expression
for multi-bunch BBU cannot be given. We can still, however, numerically
compute the individual terms equivalent to Eq.~\ref{eqypert} for the
single bunch case. For example, the first order term 
in amplitude growth is given by
\begin{equation}
\Upsilon_m= {e^2NLS_{m}\beta_0\over 2E_0} 
g(E_f/E_0,\zeta)\quad\quad\quad[m=1,\ldots,M]\ .
\end{equation}
If this term is small compared to 1 for all $m$, then BBU is well characterized
by $\Upsilon$. If it is not small, though not extremely
large, the next higher terms can be computed and their contribution added.
For $\Upsilon$ very large, this approach may not be very useful. 

From our derivation
we see that there is nothing that fundamentally distinguishes
our BBU solution from a single-bunch BBU solution. 
If we consider again the single-bunch calculation, for 
the case of a uniform charge distribution of length $\ell$, 
we see that we need to perform the integrations for $R_n$ in Eq.~\ref{eqrn}.
If we do the integrations numerically,
by dividing the integrals into discrete steps 
$t_n=(n-1)\Delta t$ and then performing
quadrature by rectangular rule, we end up with Eq.~\ref{eqcalr} with
$M=\ell/\Delta t$. The solution is the same as our multi-bunch solution. 
What distinguishes the multi-bunch from the single-bunch problem is that
the wakefield for the multi-bunch case is not normally
monotonic and does not vary smoothly with longitudinal position.
For such a case it may be more difficult to decide how many terms
are needed for the sum to converge. 

\begin{figure}[htb]
\centering
\epsfig{file=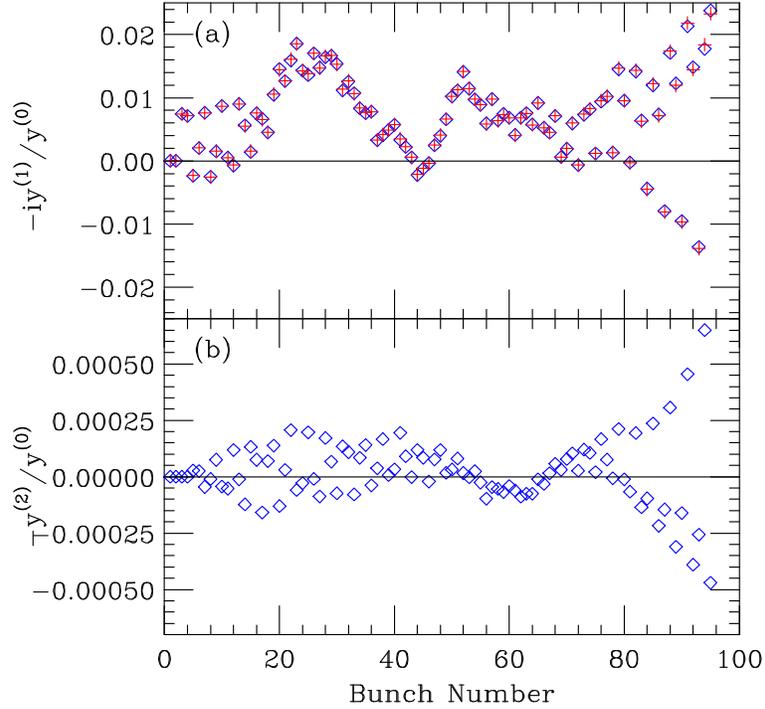, width=10cm}
\caption{
A numerical example: the NLC prelinac with the
optimized $3\pi/4$ S-band structure, but with $10^{-5}$ systematic
frequency errors, 
with the nominal (2.8~ns) bunch spacing (see Ref.~\cite{BL}).
Here $N=1.2\times10^{10}$, $E_0=1.98$~GeV, $E_f=10.$~GeV, $L=558$~m;
the rms of the sum wake $S_{rms}=.005$~MV/nC/m$^2$.
The diamonds give the first order~(a) and the second order~(b) perturbation
terms. The crosses in (a) give smooth focusing simulation results
with the free betatron term removed.
}
\label{fiperturb}
\end{figure}

In Fig.~\ref{fiperturb} we give a numerical example:
the NLC prelinac with the optimized $3\pi/4$ S-band structure,
but with $10^{-5}$ systematic
frequency errors, 
with the nominal (2.8~ns) bunch spacing (see Ref.~\cite{BL}).
The diamonds give the first order~(a) and the second order~(b) perturbation
terms. The crosses in (a) give the results of a smooth focusing
simulation program (taking $\beta\sim E^{1/2}$),
where the free betatron term has been removed.
We see that the agreement is very good; {\it i.e.} the first order term
is a good approximation to the simulation results. 
In (b) we note that the next order term is much smaller.

\section*{Acknowledgments}

The authors thanks V.~Dolgashev for
carefully reading this manuscript.

\end{document}